\documentclass[12pt,manuscript]{aastex}
\usepackage[graphicx]{}
\usepackage{multirow}

\def\gtorder{\mathrel{\raise.3ex\hbox{$>$}\mkern-14mu
             \lower0.6ex\hbox{$\sim$}}}
\def\ltorder{\mathrel{\raise.3ex\hbox{$<$}\mkern-14mu
             \lower0.6ex\hbox{$\sim$}}}
\def\proptwid{\mathrel{\raise.3ex\hbox{$\propto$}\mkern-14mu
             \lower0.6ex\hbox{$\sim$}}}
\def\0946{PG~0946+301}

\def\arcsec{\ifmmode '' \else $''$\fi}

\def\arcsecpoint{\ifmmode ''\!. \else $''\!.$\fi}

\def\kms{\ifmmode {\rm km\ s}^{-1} \else km s$^{-1}$\fi}
\def\Msun{\ifmmode {\rm M}_{\odot} \else M$_{\odot}$\fi}
\def\Lsun{\ifmmode {\rm L}_{\odot} \else L$_{\odot}$\fi}
\def\Zsun{\ifmmode {\rm Z}_{\odot} \else Z$_{\odot}$\fi}

\def\ergscm2{ergs\,s$^{-1}$\,cm$^{-2}$}
\def\icm3{{\rm cm}^{-3}}
\def\icm2{{\rm cm}^{-2}}
\def\qo{\ifmmode q_{\rm o} \else $q_{\rm o}$\fi}
\def\Ho{\ifmmode H_{\rm o} \else $H_{\rm o}$\fi}
\def\ho{\ifmmode h_{\rm o} \else $h_{\rm o}$\fi}
\def\ltsim{\raisebox{-.5ex}{$\;\stackrel{<}{\sim}\;$}}
\def\gtsim{\raisebox{-.5ex}{$\;\stackrel{>}{\sim}\;$}}
\def\vFWHM{\ifmmode v_{\mbox{\tiny FWHM}} \else
            $v_{\mbox{\tiny FWHM}}$\fi}
\def\CCF{\ifmmode F_{\it CCF} \else $F_{\it CCF}$\fi}
\def\ACF{\ifmmode F_{\it ACF} \else $F_{\it ACF}$\fi}
\def\Halpha{\ifmmode {\rm H}\alpha \else H$\alpha$\fi}
\def\Hbeta{\ifmmode {\rm H}\beta \else H$\beta$\fi}
\def\Hgamma{\ifmmode {\rm H}\gamma \else H$\gamma$\fi}
\def\Hdelta{\ifmmode {\rm H}\delta \else H$\delta$\fi}
\def\Lya{\ifmmode {\rm Ly}\alpha \else Ly$\alpha$\fi}
\def\Lyb{\ifmmode {\rm Ly}\beta \else Ly$\beta$\fi}
\def\Lyg{\ifmmode {\rm Ly}\beta \else Ly$\gamma$\fi}
\def\hi{H\,{\sc i}}

\def\hei{He\,{\sc i}}
\def\heii{He\,{\sc ii}}

\def\cii{C\,{\sc ii}}
\def\ciii{\ifmmode {\rm C}\,{\sc iii} \else C\,{\sc iii}\fi}
\def\civ{\ifmmode {\rm C}\,{\sc iv} \else C\,{\sc iv}\fi}

\def\nv{N\,{\sc v}}

\def\o5007{[O\,{\sc iii}]\,$\lambda5007$}

\def\ovi{O\,{\sc vi}}

\def\mgii{Mg\,{\sc ii}}

\def\siiv{Si\,{\sc iv}}

\def\siIII{Si\,{\sc iii}}
\def\siII{Si\,{\sc ii}}

\def\siv{S\,{\sc iv}}

\def\feii{Fe\,{\sc ii}}
\def\feiii{Fe\,{\sc iii}}

\def\alii{Al\,{\sc ii}}
\def\aliii{Al\,{\sc iii}}

\def\pv{P\,{\sc v}}

\def\o{\o}
\begin{document}

\title{BAL phosphorus abundance and evidence for immense ionic column densities in quasar outflows:\\ 
VLT X-Shooter observations of quasar SDSS J1512+1119\footnote{Based on observations collected at the European Southern Observatory, Chile, PID:87.B-0229.}}

\author{
 Benoit C.J. Borguet\altaffilmark{1},
 Doug Edmonds\altaffilmark{1},
 Nahum Arav\altaffilmark{1}, 
 Chris, Benn\altaffilmark{2},
 Carter,Chamberlain\altaffilmark{1}
}

\date{Version : \today}

\altaffiltext{1}{Department of Physics, Virginia Tech, Blacksburg, VA 24061; email: benbo@vt.edu}
\altaffiltext{2}{Isaac Newton Group, 38700 Santa Cruz de La Palma, Spain}

\begin{abstract}


We present spectroscopic analysis of the broad absorption line outflow in quasar
SDSS J1512+1119. In particular, we  focus our attention on a kinematic component
in which we identify \pv\ and \siv/\siv* absorption troughs. The shape of the
unblended phosphorus doublet troughs and the three \siv/\siv* troughs allow us
to obtain reliable column density measurements for these two ions. Photoionization
modelling using these column densities and those of \hei* constrain the
abundance of phosphorus to the range of 0.5--4 times the
solar value. The total column density, ionization parameter and metalicity inferred
from the \pv\ and \siv\ column densities leads to large optical depth values for the common
transition observed in BAL outflows. We show that the true \civ\ optical depth, is  $\sim$~1000 times greater
in the core of the absorption profile than the value deduced from it's apparent optical depth.

\end{abstract}


\section{INTRODUCTION}
\label{intro}

Active galactic Nuclei (AGN) outflows have been detected as blueshifted broad absorption lines (BALs) in the UV spectra of $\sim$ 20\% of quasars \citep{Hewett03,Dai08,Knigge08}
and as narrow absorption lines (NALs) in $\sim$ 50 \% of Seyfert galaxies \citep{Crenshaw03,Dunn08}. There is growing evidence that these ubiquitous
sub-relativistic ionized outflows play an important role on sub-parsec as well as kilo-parsec scales in controling the growth of the central black hole, the evolution
of the host galaxy and the chemical enrichment of the Intergatlactic medium (IGM) \citep[e.g.][]{Elvis06,Moe09,Ostriker10}. Moreover, the study of the abundances of metals in these outflows
observed over a range of redshifts (up to $z \sim 5$) provide us with a unique probe to investigate the history and evolution of the chemical enrichment over
cosmological scales, which constrains star formation scenarios and evolution of the host galaxy \citep{Hamann97a,Hamann98,Hamann99,Hamann03,dimatteo04,Hamann07,Germain09,Barai11}.

Studying absorption lines from AGN outflows is the most direct way to determine chemical abundances in the AGN environment. This is done by comparing the column densities
associated with ionized species of the different elements observed across the spectrum, combined with photoionization analysis. The major advantage of using absorption lines over emission lines in abundance
studies resides in the fact that they provide diagnostics that largely do not depend on temperature and density \citep{Hamann98}.  
Early abundance studies in BAL outflows implied particularly high abundances of heavy elements relative to hydrogen. In several cases, enhancement of carbon, nitrogen, oxygen and silicon by factors of tens to hundreds
of times the solar values were reported in several objects \citep[e.g.][]{Turnshek86,Turnshek96,Hamann98}, in contrast to the order of magnitude or less,
generally derived from the analysis of the quasar emission lines \citep[e.g.][]{Hamann93,Hamann02,Dietrich03,Juarez09}. 

Perhaps the most puzzling observation was the detection of BALs associated with \pv\ \citep{Junkkarinen95,Arav01,Hamann98,Hamann03}. Phosphorus is 
$\sim$ 900 times less abundant than carbon in the solar photosphere \citep{Lodders09}. Since \pv\ and \civ\ have similar ionization potentials 
they are formed in similar environments. This suggests, based on direct comparison of the measured column densities, an overabundance of phosphorus over carbon of
$\gtsim$ 100 times the solar value \citep[e.g.][]{Junkkarinen95,Junkkarinen97,Turnshek96,Hamann98}. \citet{Shields96} suggested a scenario consistent with the reported phosphorus overabundances in which the enrichment of the BAL material is mainly controlled by a population of galactic novae.
However, our group \citep{Arav97,Arav99ak,Arav99,deKool01,Arav01,Arav01b,Arav02,Arav03,Scott04,Gabel05a} and others \citep{Barlow97b,Hamann97b,Telfer98,Churchill99,Ganguly99} showed that
column densities derived from the apparent optical depth analysis of
BAL troughs are unreliable due to non-black saturation in the troughs.
Therefore, \citet{Hamann98,Hamann99,Hamann03}; and  \citet{Leighly09} suggested that
the extreme overabundance of phosphorous relative to carbon is an
artifact of very high levels of saturation in the \civ\ troughs,
compared to only mild (or non) saturation in the \pv\ troughs.
Subsequent measurements of abundances in Seyfert and quasar outflows
accounted for non-black saturation and yielded abundances of only a
few times solar in the outflows \citep{Gabel06,Arav07}.

The non-black saturation hypothesis was largely accepted by the
community to explain the \civ/\pv\ BAL observations. But this scenario
implies that the actual optical depth in the \civ\ trough is roughly
1000 times larger than the apparent one, an assertion that was never
verified empirically. In this paper, we study the UV outflow of SDSS J1512+1119, which exhibits deep absorption
troughs from \pv\ as well as \siv. In particular, we report the detection of the excited \siv\ $\lambda$1073.51 line, a transition ten times weaker than the excited \siv\ $\lambda$1072.96.
Together with estimates of the number density provided by the analysis of absorption troughs from excited states of \ciii\ and \feiii, we pinpoint the \siv\ column density.
Photoionization modeling, using the derived column densities as input, shows that the phosphorus abundance is close to the solar values.
This allows us to confirm that the true \civ\ optical depth is  $\sim$~1000 times greater in the core of the absorption profile than the value deduced from apparent optical depth measurements.

The plan of the paper is as follows: In \S~\ref{dataredu} we present the VLT/X-Shooter observations of SDSS J1512+1119 along with the reduction of the data. In \S~\ref{anabs} we
identify the spectral features and estimate the column density associated with each ionic species. We discuss the photoionization solution for the absorber and the implied phosphorus abundance in \S~\ref{discu}. We conclude the paper by summarizing the key points of the analysis in \S~\ref{conclu}.

\section{Observation and data reduction}
\label{dataredu}

SDSS J1512+1119 (J2000: RA=15 12 49.29; dec=+11 19 29.36; z=2.1062 $\pm 0.0020$ \citealt{Hewett10}; $V=17.7$), also identified as
Q1510+115 is one of the objects originally discovered in the spectroscopic
survey conducted by Hazard using objective prism plates with the UK Schmidt telescope \citep[cf.][]{sargent88}. Later spectroscopic
observations with the double spectrograph at the Palomar Hale telescope revealed the presence of broad ($\Delta v \sim 1700$ km s$^{-1}$ in \civ)
absorption troughs associated with Ly$\alpha$, \civ\, and \nv\ while several resolved components were identified in \siiv\ \citep{sargent88}.

We observed the quasar SDSS J1512+1119 with the VLT X-Shooter spectrograph on April 26 2011 as part of our program 87.B-0229 (PI: Benn).
X-Shooter is the second generation, wide band (3000 \AA ~to 24000 \AA), medium resolution ($R \sim 6000$) spectrograph installed at
the Cassegrain focus of VLT/UT2. In this instrument the incoming light is split into three independent arms, each arm consisting
of a prism-cross-dispersed echelle spectrograph optimized for the UV-blue, visible and near-IR wavelengths (UVB, VIS and NIR,
respectively) which allows for coverage of the full bandwidth in a single exposure. A detailed description of the instrument and performance
can be found in \citet{vernet11}. The total integration time for the UVB, VIS and NIR arms are 8400, 8400, and 8700 s, respectively.
The observations were performed in the slit nodding mode with two positions using a slit width of 0.8\arcsec\ in the UVB and 0.9\arcsec\
in the VIS and NIR leading to respective resolving powers $R=\Delta \lambda /\lambda$ of 6200, 8800 and 6100. 
Except for a line from the metastable 2 $^3$S excited state of \hei\ (\hei* $\lambda$3889.80, discussed in a forthcoming paper), no additional
diagnostic lines are observed within the NIR range so that we limit the current study to the UVB and VIS range of the data.

\begin{figure}
  \includegraphics[angle=90,width=1.0\textwidth]{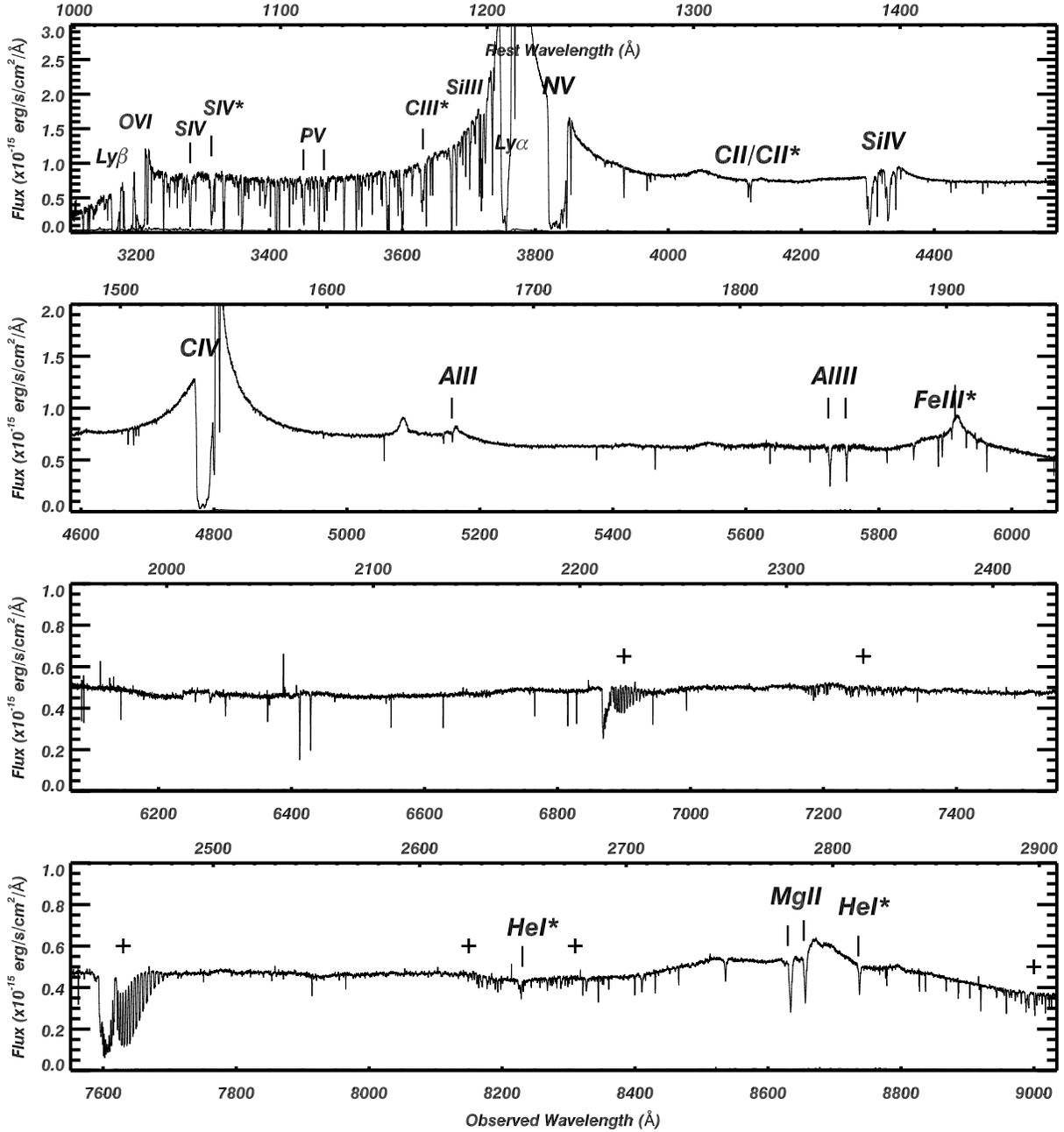}\\
 \caption{Reduced UVB+VIS X-shooter spectra of the quasar SDSS J1512+1119. We indicate the position of most of the absorption lines associated with the
intrinsic outflow (see Section~\ref{anabs}). The VIS spectrum has not been corrected for atmospheric absorption (the ``+'' indicate the location of the
major O$_2$ and H$_2$O atmospheric bands within the spectral range). This does not affect our study since
our diagnostic lines are located in regions free of such contamination.}
 \label{fulispec} 
\end{figure}

The observations were reduced in nodding mode using the ESO Reflex\footnote{Reflex is available at http://www.eso.org/sci/software/pipelines/} workflow \citep{ballester11}
and the ESO X-Shooter pipeline version 1.4.5 \citep{modigliani10} in order to obtain the rectified and wavelength calibrated two-dimensional
spectra for each arm. After manually flagging the remaining cosmic ray hits in the individual frames, we extracted one-dimensional spectra using an optimal extraction
algorithm based on the method outlined in \citet{Horne86}. An identical treatment was performed on the observation of the spectroscopic standard star
LTT7987\footnote{The flux calibrated reference spectrum can be found at: http://www.eso.org/sci/observing/tools/standards/spectra/ltt7987.html} observed
on the same day as the quasar, allowing us to flux calibrate the spectra in the range 3200 -- 9500~\AA. We present the reduced
X-shooter spectra in Figure~\ref{fulispec}.

\section{Analysis of the absorber}
\label{anabs}

\begin{figure}
  \includegraphics[angle=90,width=1.0\textwidth]{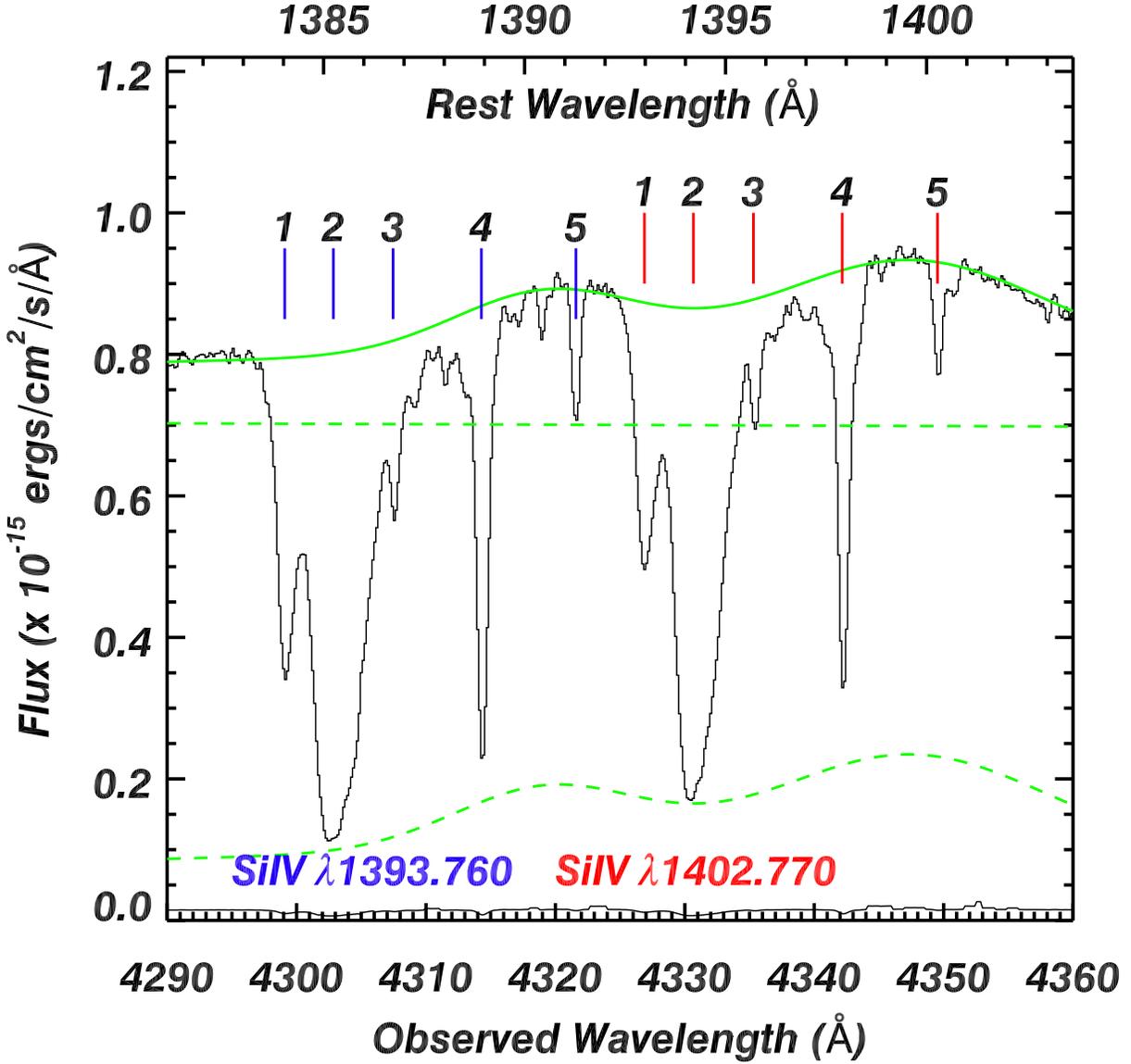}\\
 \caption{Intrinsic absorption troughs associated with \siiv\ in the VLT X-Shooter spectrum of SDSS J1512+1119.
At the resolution of the instrument, the outflow is resolved in five unblended components spanning a range of
velocities between -2100 and -520 km s$^{-1}$. The continuous line represents the unabsorbed emission model we built (see Section~\ref{coldensdet}),
and the dashed lines represents the continuum and emission line models.}
 \label{siivpro} 
\end{figure}

\begin{figure}
  \includegraphics[angle=90,width=1.0\textwidth]{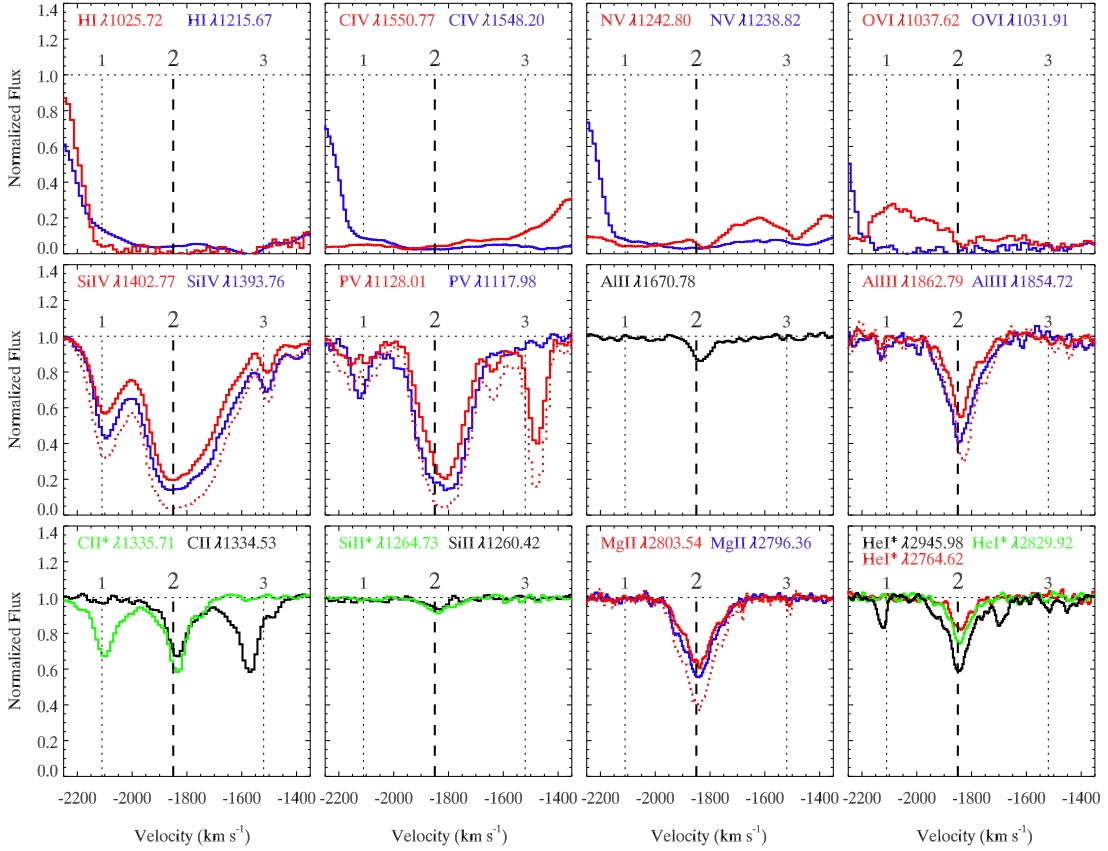}\\
 \caption{Normalized absorption line profiles of the absorption troughs associated with the
UV outflow in SDSS J1512+1119 for kinematic component 2.
For unblended doublets, we overplot the expected residual intensity in the strongest
component based on the residual flux observed in the weakest component assuming the absorbing material completely
and homogeneously covers the emission source ($I_{dotted}= I_R^2$, where $I_R$ is the residual intensity in the weaker component of the doublet).}
 \label{alllineprof} 
\end{figure}

Comparing the unblended line profile of the \siiv\ doublet, we distinguish five main kinematic components associated with the intrinsic outflow
in our VLT X-Shooter data. The centroid of the absorption features are located at radial velocities $v \sim$ -2100, -1850, -1500, -1050, and -520 km s$^{-1}$
in the rest frame of SDSS J1512+1119 (see Figure~\ref{siivpro}).
Using the \siiv\ components as a template, we identify absorption troughs associated with the kinematic components in a series of ions
spanning a range of ionization. While the kinematic components are deep and blended together in the ubiquitous Ly$\alpha$, Ly$\beta$, \ovi, \nv\ and \civ\ transitions, 
some of the components are resolved in the \mgii, \cii, \siII, \siIII, \alii, \aliii\ and \hei* lines (see Figure~\ref{alllineprof}). We report the detection
of deep absorption troughs associated with the \pv\ doublet in the component located at -1850 km s$^{-1}$ (component 2, see Figure~\ref{alllineprof}), as well as 
absorption associated with ground state and excited \siv, and excited \ciii\ and \feiii. Of particular interest in this component is the detection of the high ionization excited \siv\ $\lambda$1073.51 
transition, a line ten times weaker than the \siv\ $\lambda$1072.96 arising from the same excited level. The combined detection of these lines allows us to accurately determine the total
\siv\ column density for that system (see Section~\ref{sivcoldens}). In the remainder of this analysis we will focus our attention on
component 2, deferring the study of the other components to a forthcoming paper.

\subsection{Ionic column densities}
\label{coldensdet}

In order to compute the column densities, we first need to estimate the spectrum of the background source (continuum and emission lines) illuminating the absorbing material.
We model the unabsorbed and dereddened (E(B-V)=0.051 \cite{Schlegel98}) continuum by a single power law of the form $F_{\lambda}=F_{1100}(\lambda/1100)^{\alpha}$, where $F_{1100}=77.01 \pm 0.05 \times 10^{17}$
erg/cm$^2$/s/\AA\ is the observed flux at 1100 \AA\ (rest frame) and the power law index $\alpha = -0.403 \pm 0.004$. 
Given the overall high S/N of the X-Shooter data (S/N $\sim$ 30 -- 70 over most of the UVB/VIS range), we then fit the unabsorbed
line emission using a smooth third order spline fit. We constrain the emission model over the broader troughs associated with Ly$\beta$,
\ovi, Ly$\alpha$, \nv, and \civ\ by using the profiles derived from the unabsorbed red wings of each emission line. This fitting procedure
is particularly effective given the tight constraints on the unabsorbed emission due to the high S/N and the presence of mainly narrow
absorption features across the spectrum.

Once normalized by the emission profile, we extract the ionic column densities $N[$ion$]$ for each species by
modeling the observed residual intensity $I_i$ (see Figure~\ref{alllineprof}) inside the troughs as a function of the radial velocity. We use three different models in order to account for
inhomogeneities in the absorbing material; apparent optical depth (AOD), partial covering (PC) and a power law
model (PL) \citep[see][for details]{Edmonds11,Borguet12}. We integrate the line of sight averaged column density \citep{Hamann04,Edmonds11} over the radial velocity range $v \in [-2030,-1570]$ km s$^{-1}$
corresponding to kinematic component 2 and report the values in Table~\ref{coldensi}. We only report lower limits derived with the AOD model on column densities
for singlet lines (\alii, \cii, \siII) as well as for deep/self-blended troughs such as \ovi, \civ, and \nv.

For photoionization modeling, actual $N[$ion$]$ measurements with robust error bars are much more useful
than lower (or upper) limits. We now demonstrate that the troughs of \aliii, \pv\ and \hei* are not
heavily saturated and therefore reliable $N[$ion$]$ measurements can be extracted for them. In Section~\ref{sivcoldens}
we show that this is also the case for \siv*. Let us first examine the \aliii, and \pv\ doublets, where
the oscillator strength $(f)$ of the blue transition is twice that of the red transition. From
Figure~\ref{alllineprof} we can infer that in both cases the residual intensity of the red doublet
component ($I_r$) is significantly higher than residual intensity of the blue doublet component ($I_b$)
along essentially the entire absorption trough. Such a situation implies the true ionic column density in
the trough cannot be much larger than 2-3 times that of an AOD estimate (see detailed treatment of this issue
in \citealt{Arav05,Arav08}). Our Table~\ref{coldensi} shows that this is indeed the case for \aliii, and
\pv, where we are able to measure robust $N[$ion$]$ for these ions using all three absorption models.
If we define the saturation as $S \equiv N[$ion$]$(PL)/$N[$ion$]$(AOD) then the level of saturation
in these troughs (from Table~\ref{coldensi}) is less than 1.2 for \aliii, and less than 3 for \pv\ (even
when taking the large upper error bar on the \pv\ measurement). Similarly the \hei* is even less saturated ($S<1.1$).
Under these conditions, the exact behavior of the covering fraction as a function of velocity ($C(v)$) across the
trough for the PC model, or that of the exponent $a(v)$  of the PL method, do not affect the robustness of the
derived $N[$ion$]$. 

While \siiv\ does not seem particularly saturated in Figure~\ref{alllineprof}, we have to stress the fact that
the normalization of the line profile was performed under the assumption that the absorbing material covers the emission sources (continuum and emission lines)
by the same fraction. A closer inspection of the non-normalized \ovi\ and \civ\ absorption troughs, however, reveal the existence of residual emission indicating
that at least part of the emission lines (the intermediate/narrow emission component) are not fully covered by the absorber as observed in other outflows \citep[e.g.][]{Arav99,Gabel06}.
Failure to account for this observation leads to an underestimation of the total \siiv\ column density since the weaker line emission under the blue absorption trough relative to the
red would mislead the observer by making the doublet appear non-saturated (see Figure~\ref{siivpro}). For this reason, we report a lower limit on the \siiv\ column density in Table~\ref{coldensi}.
Another transition that could be affected by this problem is \mgii, although in this case, the weaker underlying line emission would only affect the absorption profile of the
red component of the doublet. Other non-saturated species (\pv, \aliii\ etc.) are located further away from significant emission lines in the SDSS J1512+1119 spectrum and
are therefore not affected by this problem. In the last column of Table~\ref{coldensi}, we report the
column densities used in the photoionization analysis. Those column densities are selected according to the following procedure: we use the values
reported in the PC column as the measurements and the PL measurement and error as the upper error in order to account for possible inhomogeneities
in the absorber \citep[see][]{Borguet12}. When only AOD determinations are available, we consider the reported values minus the error as a lower limit.

 \begin{deluxetable}{lrrrr}
\tablecaption{{\sc Computed column densities}}
\tablewidth{0pt}
\tablehead{
\colhead{Ion}
&\colhead{AOD$^{\mathrm{a}}$}
&\colhead{PC$^{\mathrm{a}}$}
&\colhead{PL$^{\mathrm{a}}$}
&\colhead{Adopted$^{\mathrm{b}}$}\\
\colhead{}
&\colhead{$10^{12} \mathrm{cm}^{-2}$}

&\colhead{$10^{12} \mathrm{cm}^{-2}$}

&\colhead{$10^{12} \mathrm{cm}^{-2}$}

&\colhead{$10^{12} \mathrm{cm}^{-2}$}

}
\startdata

   \hi     &  $> 765$                 &       ...                         &                         &   $> 765$     \\
   \hei*    &  687$^{+22}_{-21}$       &   695$^{+10}_{-10}$               &  715$^{+27}_{-9}$       &   695$^{+47}_{-10}$     \\
   \cii    &  214$^{+4}_{-4}$         &       ...                         &                         &   $> 210$     \\
   \civ    &  $> 2550$                &       ...                         &   ...                   &   $> 2550$     \\
   \nv     &  $> 3280$                &       ...                         &   ...                   &   $> 3280$      \\
  \ovi     &  $> 3940$                &       ...                         &   ...                   &   $> 3940$      \\
   \mgii   &  24.8$^{+0.3}_{-0.3}$    &         35.6$^{+0.6}_{-0.5}$      &   51.9$^{+1.0}_{-1.0}$  &   35.6$^{+17.3}_{-0.5}$     \\  
   \alii   &  1.76$^{+0.08}_{-0.08}$  &       ...                         &                         &   $> 1.68$    \\  
   \aliii  &  42.4$^{+0.7}_{-0.7}$    &         51.0$^{+0.8}_{-0.7}$      &   55.7$^{+0.8}_{-0.7}$  &   51.0$^{+5.5}_{-0.7}$     \\  
   \siII   &  4.05$^{+0.37}_{-0.35}$  &       ...                         &   ...                   &   $>3.7$     \\  
   \siIII  &  23.3$^{+0.2}_{-0.2}$    &       ...                         &   ...                   &   $> 23.1$     \\  
   \siiv   &  391$^{+2}_{-2}$         & 529$^{+5}_{-4}$                   &   $> 800$               &   $>$525      \\  
   \pv     &  343$^{+5}_{-5}$         &             444$^{+13}_{-10}$     &  695$^{+211}_{-8}$      &   444$^{+462}_{-10}$    \\  
   \siv$^{\mathrm{c}}$    &  ...                     &    ....                           &  ...                    &  26600$^{+4000}_{-2700}$    \\

\enddata
\label{coldensi}
a) The integrated column densities for the three absorber models. The quoted error arise from photon statistics only and are computed using the technique outlined in~\citet{Gabel05a}.\\
b) Adopted values for the photoionization study (see text).\\
c) See Section~\ref{sivcoldens} for details.
\end{deluxetable}

\subsection{The \siv\ column density}
\label{sivcoldens}

The combined good S/N, medium resolution and moderate Ly forest contamination in the X-Shooter
spectrum of SDSS J1512+1119 allow us to identify and separate the absorption troughs associated with \siv\ in kinematic component 2 of
the outflow. As discussed in \citet{Leighly09,Leighly11}, \siv\ is composed of three lines; the ground state transition with a
wavelength of 1062.66 \AA, and two transitions arising from an excited state ($E=951$ cm$^{-1}$, referred to as \siv* in the following) at wavelengths 1072.96 and
1073.51 \AA, the excited state being populated at a critical density of $4.7 \times 10^4$ cm$^{-3}$. A useful feature of
these lines is that the fractional abundance of \siv\ peaks at a similar ionization parameter to that of the ubiquitous \civ, implying that
they arise in similar regions of the outflow \citep{Dunn12}. The presence
of two excited \siv* transitions with oscillator strengths an order of magnitude apart \citep{Hibbert02} provides sensitivity to a wide
range of optical depths before both troughs become fully saturated.

\begin{figure}
  \includegraphics[angle=90,width=1.0\textwidth]{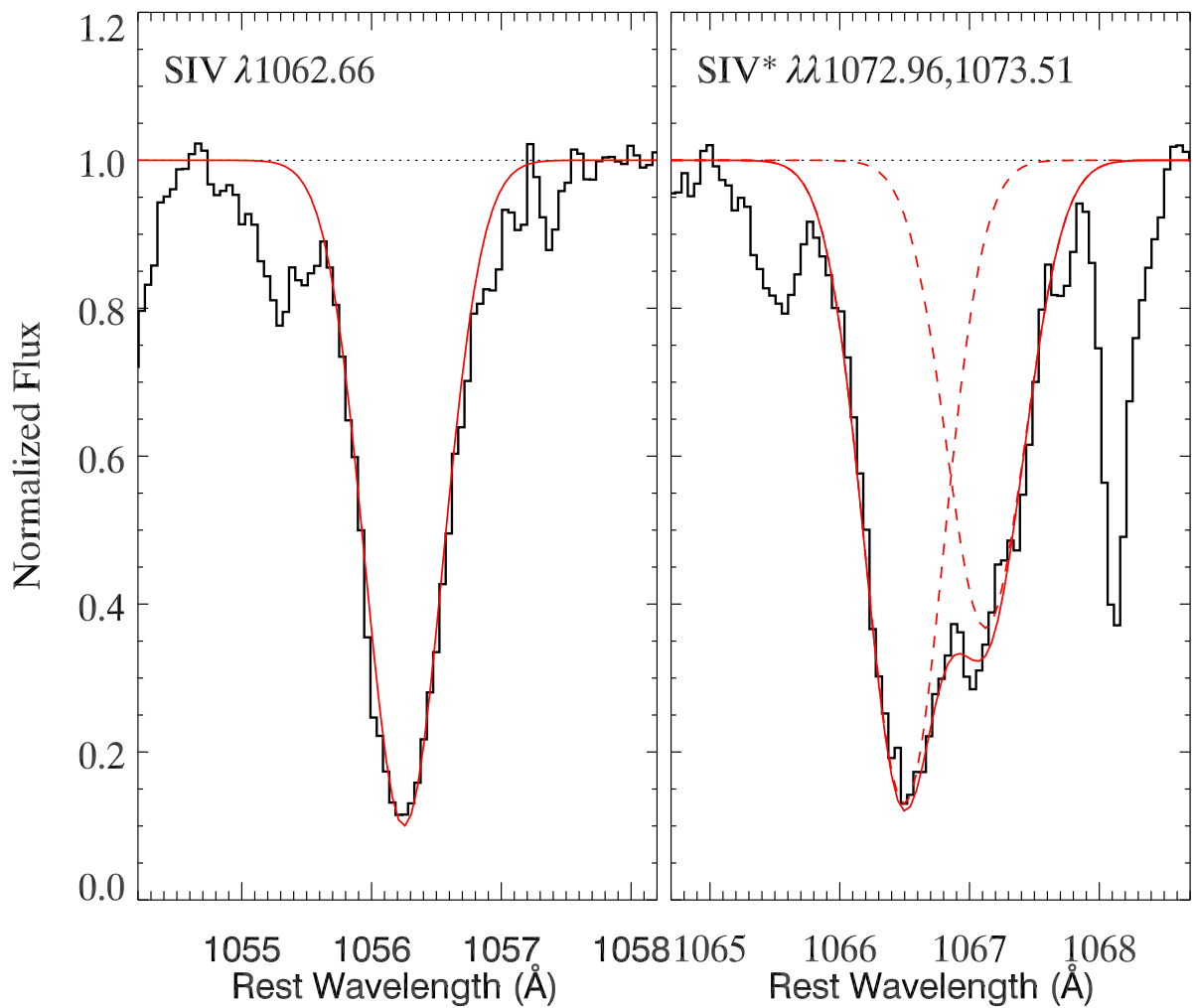}\\
 \caption{Result of the simultaneous fit (continuous smooth line) to the three \siv\ lines present in component 2.
Within the uncertainties, the fitted profiles of the resonance $\lambda$1062.66 and excited $\lambda$1072.97 lines are identical, suggesting that
the line profiles are saturated and no column density information can be derived from them. However, the detection of the excited 1073.51 \AA\
line, a transition ten times weaker than the 1072.97 \AA\ line, allows us to estimate the real \siv* column density.}
 \label{sivsivefit} 
\end{figure}

In Figure~\ref{sivsivefit}, we present the result of the simultaneous Gaussian fit performed over the observed line profiles of the three \siv\ lines associated with
component 2 of the outflow. The fit was performed by fixing the wavelength positions of the lines relative to the velocity of the component
derived from the strong \pv\ $\lambda 1117.98$ line profile as well as imposing an identical $FWHM$ for the three Gaussians, leaving
only the depth of each component as a free parameter. The result of the fit clearly shows a detection of the weak excited $\lambda$1073.51 transition,
translating to a high \siv* column density. Such a high column density leads to the conclusion that the $\lambda$1072.97
transition is optically thick and that the non-black profile observed is due to the partial covering of the emission
source by the absorbing material. The depth of the resonance $\lambda$1062.66 line is consistent, within the uncertainties, with the depth derived
from the $\lambda$1072.97 excited line implying that the $\lambda$1062.66 profile is also only reflecting a partial covering of the emission source.
Note that the presence of a trough associated with component 1 of the outflow is observed in \siv\ around 1055.2 \AA\ and in \siv* $\lambda$1072.97 around 1065.5 \AA\
could affect the result of the fit if the optical depth in that system is large enough to produce a significant \siv*  $\lambda$1073.51 trough.
However, the high covering deduced from other high ionization lines (e.g. \civ, \siiv) for that component suggests a low optical depth ($\tau \simeq 0.2$) associated
with the $\lambda$1072.97 translating to a $\tau \simeq 0.02$ for the ten times weaker $\lambda$1073.51 transition. Such a tiny optical depth
will not affect the presented modeling of trough 2.
Using the Gaussian model of the non-saturated $\lambda$1073.51 line profile along with the fact that the $\lambda$1072.97 line profile is saturated
provides an unequivocal determination of the covering fraction across the trough (i.e. $C=1-I_i$ for saturated lines). We solve the PC model residual
intensity equations for both \siv* transitions simultaneously and estimate the total \siv* column density to be $N[$\siv*$]\sim 1.6 \times 10^{16}$ cm$^{-2}$.

\begin{figure}
  \includegraphics[angle=90,width=0.95\textwidth]{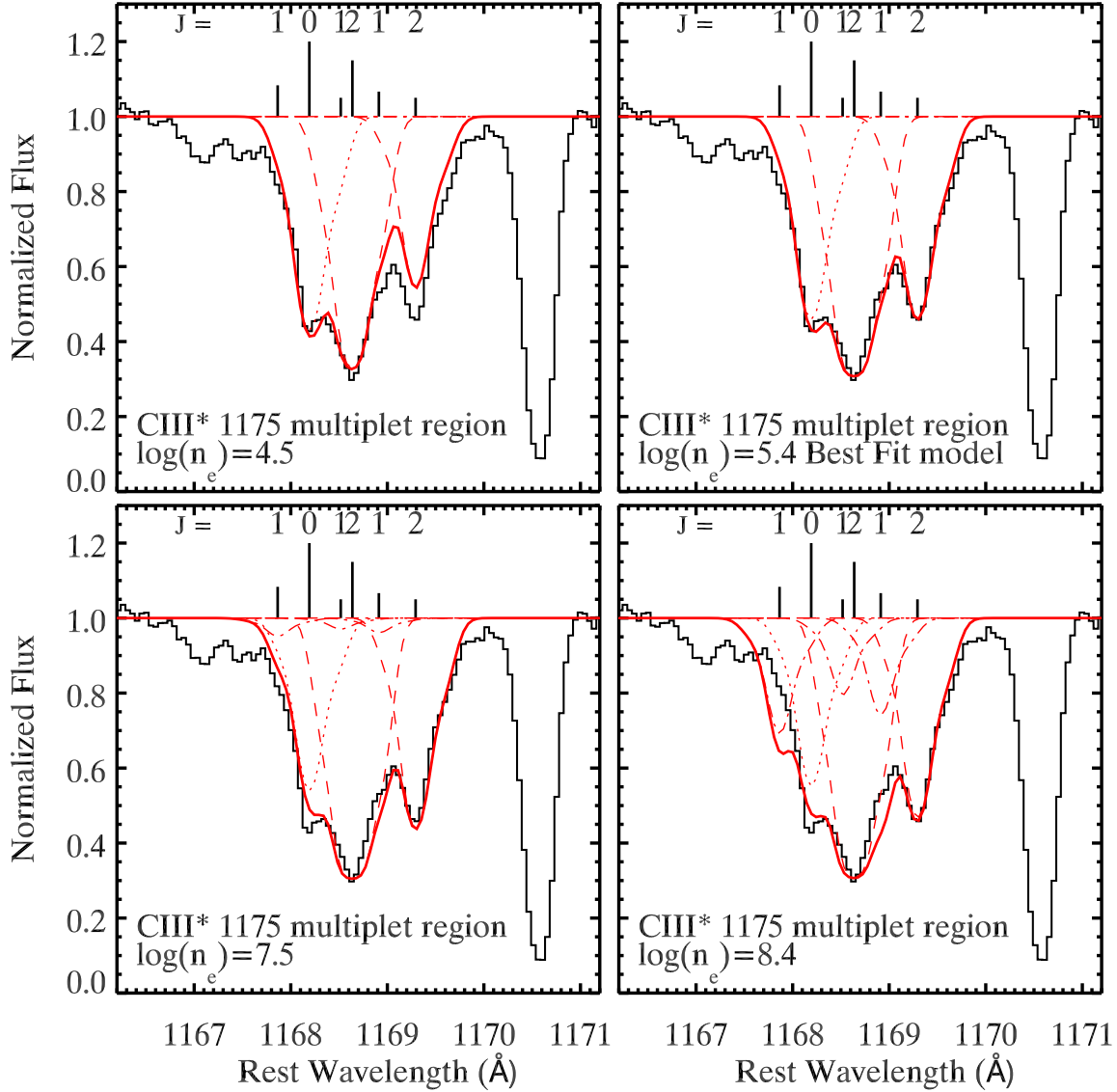}
 \caption{Fits to the normalized \ciii* multiplet blend (continuous line) for component 2. The vertical lines plotted under each sub-component
of the multiplet shows the location of the centroid of the line and the length of the mark indicates the relative oscillator
strength of the line. The absorption line located at 1170.6 \AA~ is an unrelated line belonging to the Ly forest.
 We present the fit of the blended profile assuming the ratio between the J levels for four representative electron number densities.
The lower limit on the possible $n_e$ is constrained by the inability of the model to fit the absorption associated with the longer wavelength J=2 transition
(upper left). The upper limit
on the $n_e$ range is fixed by the the apparition of significant absorption associated with the J=1 level for $\log(n_e) \geq 8.1$ (lower right).
The best fit of the blend is obtained for $\log(n_e)=5.1$ (upper right panel). Other densities within the range $4.8 \geq \log(n_e) \geq 8.1$
lead to acceptable fits of the blend (lower left panel).}
 \label{ciiifit} 
\end{figure}

In order to determine the total column density in \siv, we also have to be able to estimate the column density present in the
ground state transition. As detailed above, the $\lambda$1062.66 line is saturated so that no accurate column
density can be derived from the depth of the line profile. However, knowing the electron density $n_e$ of the gas would
allow us to estimate the total \siv\ column density by comparing the measured value to models predicting the population ratio in the excited state to the ground state as a
function of $n_e$. The detection of a blend of lines that we identify with the \ciii* multiplet gives us the possibility to do so. As detailed in \citet{gabel05b},
the excited \ciii\ $\lambda$1175 multiplet comprises 6 lines arising from 3 J levels. The J=0 and J=2 levels have significantly lower radiative transition probabilities
to the ground state than the J=1 level and are thus populated at much lower densities than the latter. In particular, Figure~5 in \citet{gabel05b} shows that the
relative populations of the three levels are a sensitive probe to a wide range of $n_e$ while being insensitive to temperature. In Figure~\ref{ciiifit},
we show several fits of the \ciii* multiplet assuming a broadened Gaussian optical depth distribution for each line. The broadened Gaussian
optical depth profile is identical for each line of the blend, only the optical depth varyies between the lines. The profile composed of a main
central Gaussian profile (FWHM $= 70$ km s$^{-1}$) containing the
core of the optical depth ($\tau_{core}$) centered at the radial velocity identical to the centroid of the \pv\ $\lambda 1117.98$ feature. Two weaker Gaussians ($\tau = 0.2 \tau_{core}$)
with FWHM $= 70$ km s$^{-1}$ are added at $\pm 70$ km s$^{-1}$ of the core Gaussian in order to produce the broadened wings. The fitting model is tightly constrained
by the exact kinematic position (derived from the \pv\ trough) and given the kinematic separation of the six \ciii* components. 
The fit was repeated for various electron densities in the range $\log (n_e) \in [3,10]$ with a step of 0.1 dex. The only free parameters for each model were a single optical
depth and covering fraction. The best fit was found using a $\chi^2$ figure of merit for a density of $\log (n_e) \sim 5.4$ cm$^{-3}$.
We estimated conservative error-bars on the electron density by noting that for densities lower than
$\log (n_e) \gtsim 4.8$ the population of the J=2 level is too low to produce enough absorption (upper left panel of Figure~\ref{ciiifit}) while the absence of
significant contribution from J=1 lines limits the maximum density to $\log (n_e) \ltsim 8.1$ (lower right panel of Figure~\ref{ciiifit}). Electron number
densities within the range $4.8 \geq \log(n_e) \geq 8.1$ provides acceptable fit to the blend.
The density estimated from the \ciii* fit is in good agreement with the detection of \feiii* lines from the UV34 multiplet (see Borguet et al. 2012b, in
preparation) which have a critical density of $\sim ~10^5$ cm$^{-3}$ and with the absence of \feiii* lines from the UV48 multiplet for which the critical
density is $\sim ~10^{10}$ cm$^{-3}$ (Bautista 2008, private communication). Assuming the electron density derived above and that
the level population of \siv* relative to \siv\ are determined by collisional excitation and radiative de-excitation only, we are able to estimate the 
column density in the \siv\ ground state to be $1.06^{+0.40}_{-0.27}$ $10^{16}$ cm$^{-2}$, where the errors reflect the uncertainty on $n_e$.

\section{Photoionization modeling and relative abundance of phosphorus}
\label{discu}

\subsection{Physical state of the gas assuming solar abundances}
\label{physstate}

In this section we use the column densities determined for each of the ionic species to constrain
the physical state (ionization parameter $U$, and total hydrogen column density $N_H$) of the gas. We model the absorber by a plane parallel
slab of gas of constant hydrogen number density ($n_H$) and use the spectral synthesis code Cloudy C10.00 (last described in \citealt{Ferland98})
to solve the ionization equilibrium equations. We assume solar abundances for the gas from \citet{Lodders09}. Due to the lack of constraints on the EUV/FUV region of the spectral energy
distribution (SED) of SDSS J1512+1119, we adopt the ``UV-soft'' SED proposed for luminous, Radio Quiet quasars in \citet{Dunn10a}; their Figure 11.
The features of this SED, which is discussed in detail in \citet{Dunn10a}, departs from the ``classical'' \citet{Mathews87} SED (MF87) by excluding the so called
UV-bump peaking at FUV energies while keeping an $\alpha_{ox}$ index similar to that of MF87.   

We use the grid model approach described in \citet{Edmonds11} in order to determine the pair of parameters ($U$,$N_H$) that
best reproduces the observed ionic column densities. We only consider column densities determined from non-saturated troughs, since the
lower limits placed on the column densities from \hi, \civ, \nv, \ovi\ are clear underestimations of their true values (see Section~\ref{coldensdet}) and are consistent with the
solution derived from the non-saturated lines. We have constraints from 10 ionic species : \hei*, \cii, \mgii, \alii,
\aliii, \siII, \siIII, \siiv, \pv\ and \siv. From this set, \pv, \siv, \hei* and \aliii\ are the most reliable as their
$N[$ion$]$ is derived from two or more troughs that are not heavily saturated (see Section~\ref{coldensdet}).

 Given that the derived density of the absorbing material is well over the \cii\ critical density we conclude that our estimation of the \cii\ column density
is robust within a factor of two, since for such a high density, the apparent strength of \cii* $\lambda 1335.71$ is approximately twice that of
\cii\ $\lambda 1334.53$, allowing us to derive a PC and PL solution by applying a similar treatment to these lines as to resonance doublets.
The other lines are either singlets for which a robust $N[$ion$]$ is difficult to ascertain in principle, or doublets that may be more heavily saturated (\siiv, \mgii, see Section~\ref{coldensdet}). We finally place an  upper limit on the column density of the non-detected \feii\
by scaling the \mgii\ blue line profile template to the $2 \sigma$ noise in the region where the strongest \feii\ $\lambda$2382.77
should be located and find $N[$\feii$]< 1.96 \times 10^{12}$ cm$^{-2}$.

\begin{figure}
  \includegraphics[angle=0,width=0.8\textwidth]{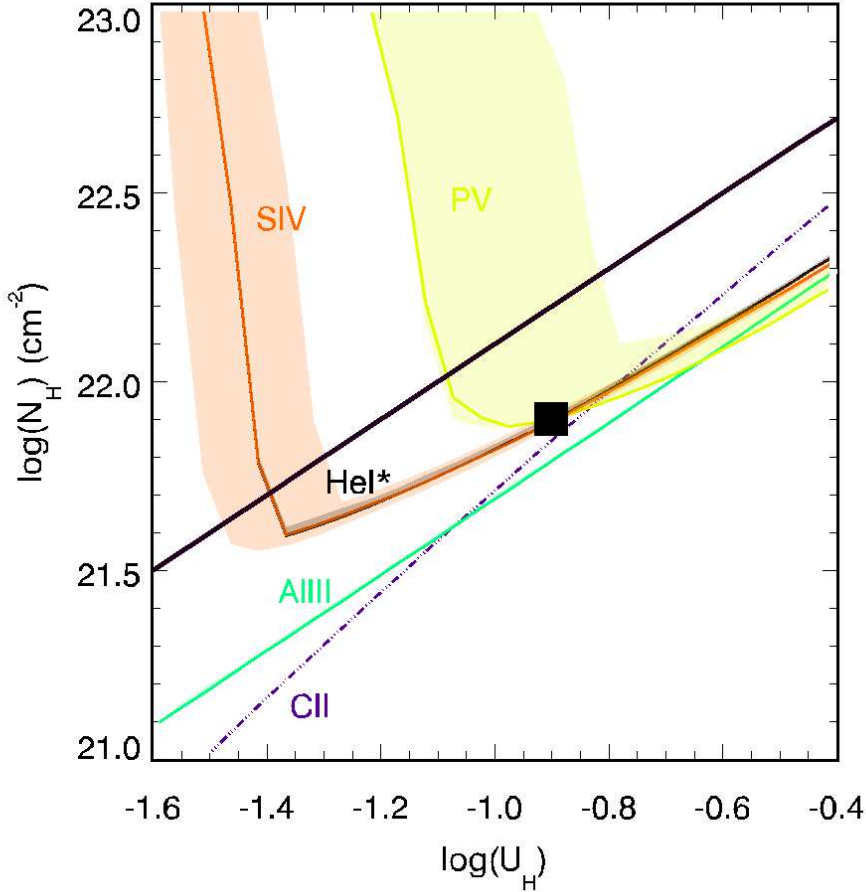}\\
 \caption{Grid model plot of the photoionization modeling of the absorber. Solid lines are the adopted $N[$ion$]$ measurements
from Table~\ref{coldensi}. The shaded area represent the  measured column density of the ion
and its errors on these measurements due to the photon noise and the systematic uncertainties in the absorber model. Dotted-dashed line represent a
lower limit on the column density. The black square represents the $U$ and 
$N_H$ that best reproduces the observed column densities (see text). The upper part of the phase plot (beyond the hydrogen ionization front characterized by
$\log(N_H) = 23.1 ~\log(U)$) is an artifact of our modeling procedure. By coincidence the \hei* and \siv\ lines almost track each other between $\log(U)=-1.4$ and $\log(U)=-0.5$.
We represent the hydrogen ionization front ($\log(N_H) = 23.1 \log(U_H)$) by a thick black line.}
 \label{gridmodel} 
\end{figure}

%
%
%
%
%
%
%
%
%
%
%
%
%
%
%
%
%
%
%
%
%

\begin{deluxetable}{lccc}

 \tablewidth{0.6\textwidth}

 \tablecolumns{4}

 \tabletypesize{\footnotesize}

 \tablecaption{Photoionization model of the absorber}

 \tablehead{

 \colhead{Ion} &

 \colhead{log($N_{obs}$) (cm$^{-2}$)} &

 \colhead{log($N_{mod}$) (cm$^{-2}$)} &
\colhead{log$\left(\frac{N_{mod}}{N_{obs}}\right)$}\\

 \colhead{} &

 \colhead{Adopted $^{\mathrm{a}}$} &

 \colhead{Cloudy} &
 \colhead{} 
  }

 \startdata

 log $(U)$ & $\cdots$ & -0.90$^{\mathrm{b}}$ & \\

 log $(N_H)$ & $\cdots$ & 21.9$^{\mathrm{b}}$ & \\

\hline

 \hi\   &  $> 14.88$             & 17.51 & 2.63 \\
\bf{ \hei*\ } & $14.84^{+0.03}_{-0.01}$ & 14.86 & \bf{ 0.02 } \\
\bf{ \cii\ } &  $\sim 14.33$              & 14.50 & \bf{ 0.19 } \\
 \civ\  &  $>15.41$              & 17.91 & 2.50 \\
 \nv\   &  $>15.52$              & 16.91 & 1.40 \\
 \ovi\  &  $>15.60$              & 17.81 & 2.21 \\
 \mgii\ &  $13.55^{+0.17}_{-0.01}$ & 14.34 & 0.79 \\
 \alii\ &  $>12.23$              & 12.45 & 0.22 \\
\bf{ \aliii\ } &  $13.71^{+0.05}_{-0.01}$ & 14.13 & \bf{ 0.42 }\\
\siII\  &  $>12.57$              & 12.96 & 0.39 \\
\siIII\ &  $>13.36$              & 15.09 & 1.73 \\
\siiv\  &  $>14.71$              & 16.24 & 1.52 \\
\bf{ \pv\ }   &  $14.65^{+0.31}_{-0.01}$ & 14.65 & \bf{ 0.00 }\\
\bf{ \siv\ }  &  $16.43^{+0.06}_{-0.04}$ & 16.45 & \bf{ 0.02 }\\

 \enddata


 \label{cloudy_model}
 a) Adopted column densities reported in Table~\ref{coldensi}. Ions with robust measurements
are marked in boldface.\\
b) Best fit Cloudy model.
\end{deluxetable} 

We present the results for a grid of photoionization models in Figure~\ref{gridmodel}. Visual inspection of the figure
shows that a minimum ionization parameter of $\log(U) > -1.1$  is required by the \pv\ constraints. The presence of the low ionization
\cii, \siII, and \alii\ requires  $\log(U) < -0.6$: although these low ionization species can also be formed at higher $U$ near
a hydrogen ionization front in the absorber, such a model would overpredict well constrained ions like \siv\ or \pv\ by a factor $\geq 5$. The corresponding total hydrogen
column density is rather large due to the detection of \pv\ as well as the high column density
derived from the detailed analysis  of the \siv\ troughs and suggests a thick absorber with  $\log(N_H) \in [21.5,21.9]$ for the
range of ionization parameters considered here. Such a high column density is also supported by the detection of \hei*, which implies an absorber with a
thickness reaching well into the \heii\ region, while the clear non-detection of the strongest \feii\ lines implies the absence
of the formation of a hydrogen ionization front in the slab.
We determine the best ($U,N_H$) model by $\chi^2$ minimization of the difference between the measured column densities
and those predicted by Cloudy \citep[see][]{Borguet12} and find $\log(U) =-0.9$ and $\log(N_H) =21.9$. In Table~\ref{cloudy_model},
we compare the measured column densities to those predicted by the best-fit model. Taking the uncertainties into account, the 
column densities associated with the ions for which we have a robust measurement are matched well within a factor of 2 while the \aliii\ column density is reproduced 
within a factor of 3. The computation shows also an underestimation by factors of hundreds for the usual high ionization
lines, underlining the fact that AOD measurements yield a poor estimate of the actual column densities for these species.

\subsection{Constraining the relative phosphorus abundance}
\label{phosab}

In this section, we use our knowledge of the photoionization solution and the measurements of the column densities of \pv\ and mainly \hei* in order to constrain the abundance of phosphorus
in the outflowing material.
Let us first assume that all elements except phosphorus have solar abundances. Given the photoionization solution determined in Section~\ref{physstate}, the abundance of phosphorus 
relative to helium is constrained by the upper-limit on the column density of \pv\ to be $\ltsim 2$ times the solar value. In these models, we use a hydrogen number density of $10^4$~cm$^{-3}$, which is close to the lower limit determined in Section~\ref{sivcoldens}. Increasing the hydrogen number density to $10^8$~cm$^{-3}$, the upper limit on the number density, increases $U$ and $N_H$ by 0.1 dex each and reduces the upper limit on the abundance ratio P/P$_\odot$. We therefore conclude that, for solar abundances of the other elements, the phosphorus abundance is approximately solar.

\begin{figure}
  \includegraphics[angle=0,width=0.8\textwidth]{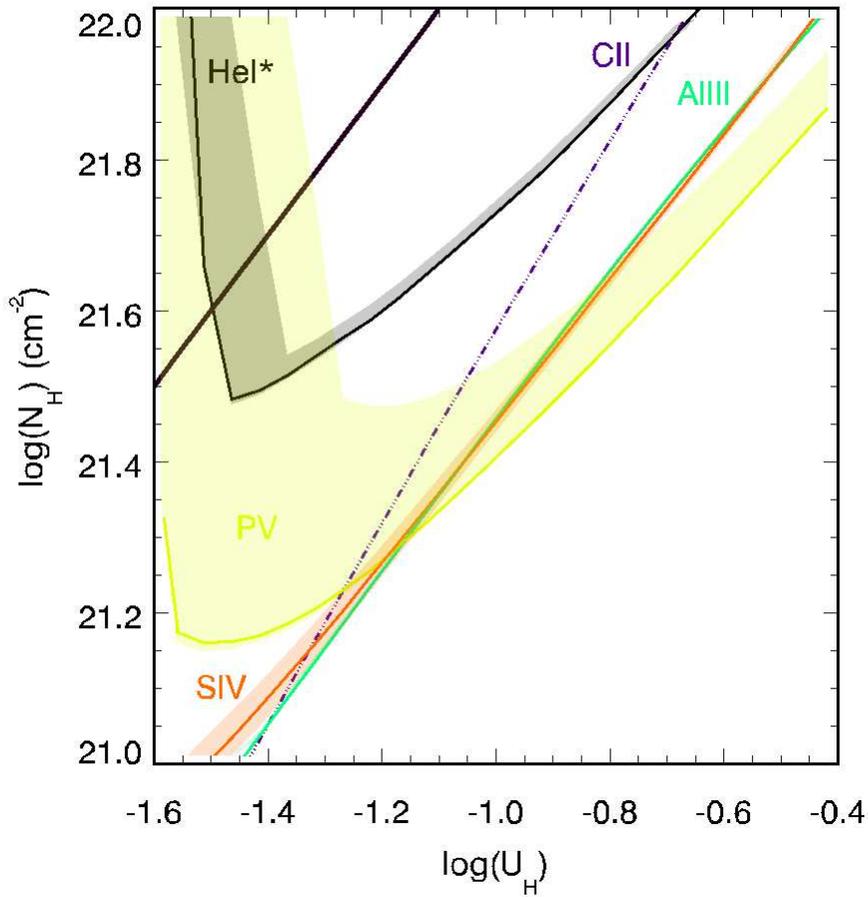}\\
 \caption{Identical to Figure~\ref{gridmodel} but assuming a metalicity of $Z=4$ (see Section~\ref{phosab} for details). It is clear that
no $U$ and $N_H$ solution can simultaneously satisfy the \siv\ and \hei* measurements, as these curves never intersects.}
 \label{gridmodel2} 
\end{figure}

While the parameter fit using solar abundances is very good, we consider the effects of changing metalicity on the phosphorus abundance. For our purposes, we consider a gas with metalicity Z/Z$_\odot$~=~1 to have solar abundances, noting that the scaling of heavy elements with Z is model dependent \citep[e.g.][]{Hamann93,Korista96}. Figure~\ref{gridmodel2} is a grid model plot for photoionization models with Z/Z$_\odot \approx 4$. We use the abundance scalings provided in Table~2 of \citet{Ballero08} for C, N, O, Mg, Si, Ca, and Fe. Scalings for the other metals as well as helium are estimated using starburst models in Cloudy for Z/Z$_\odot$~=~4. For these models, any solution that reasonably reproduces column densities of the metals underestimates the column density of helium by a factor $\sim 2$ for a number density of $10^4$~cm$^{-3}$ and less than a factor of 2 for a number density of $10^8$~cm$^{-3}$ . Increasing the metalicity increases this discrepancy, thus we conclude that Z/Z$_\odot \ltsim 4$. We find that none of the models accurately reproduce our measured ionic column densities if the metal abundances are reduced by a factor $\sim 2$, implying that the gas is has a metalicity Z/Z$_\odot \gtsim 0.5$. Constraining the metalicity to 0.5$\le$Z/Z$_\odot <$4 limits the ionization parameter to $-1.3 \ltsim U \ltsim -0.5$. Comparing phosphorus only to helium, we find that $1/2 < $P/P$_\odot < 5$. The upper limit is overly conservative and is the maximum over-abundance of phosphorus that allows the models to produce enough \hei*.

Ionization and thermal structures in the absorber depend on the SED incident on the outflowing gas. In the foregoing analysis, we used the UV-soft SED mentioned in Section~\ref{physstate}. We tested several SEDs appropriate for radio-quiet quasars to determine their impact on our main results. The largest changes occurred for SEDs including a substantial ``UV bump''. In particular, using MF87, we find the best fit model with solar abundances yields log~$U = -1.1$ and log~$N_H = 21.6$. However, more importantly for our purposes here, our diagnostic line ratios change by factors $\ltsim 2$ for corresponding models. The major change is that the conservative upper limit on P/P$_\odot$ can be as large as 6 for the MF87 SED. However, we emphasize again that the MF87 SED is not a good approximation to a high luminosity radio quiet quasar.

\subsection{The true \civ\ optical depth}

\begin{figure}
  \includegraphics[angle=90,width=1.0\textwidth]{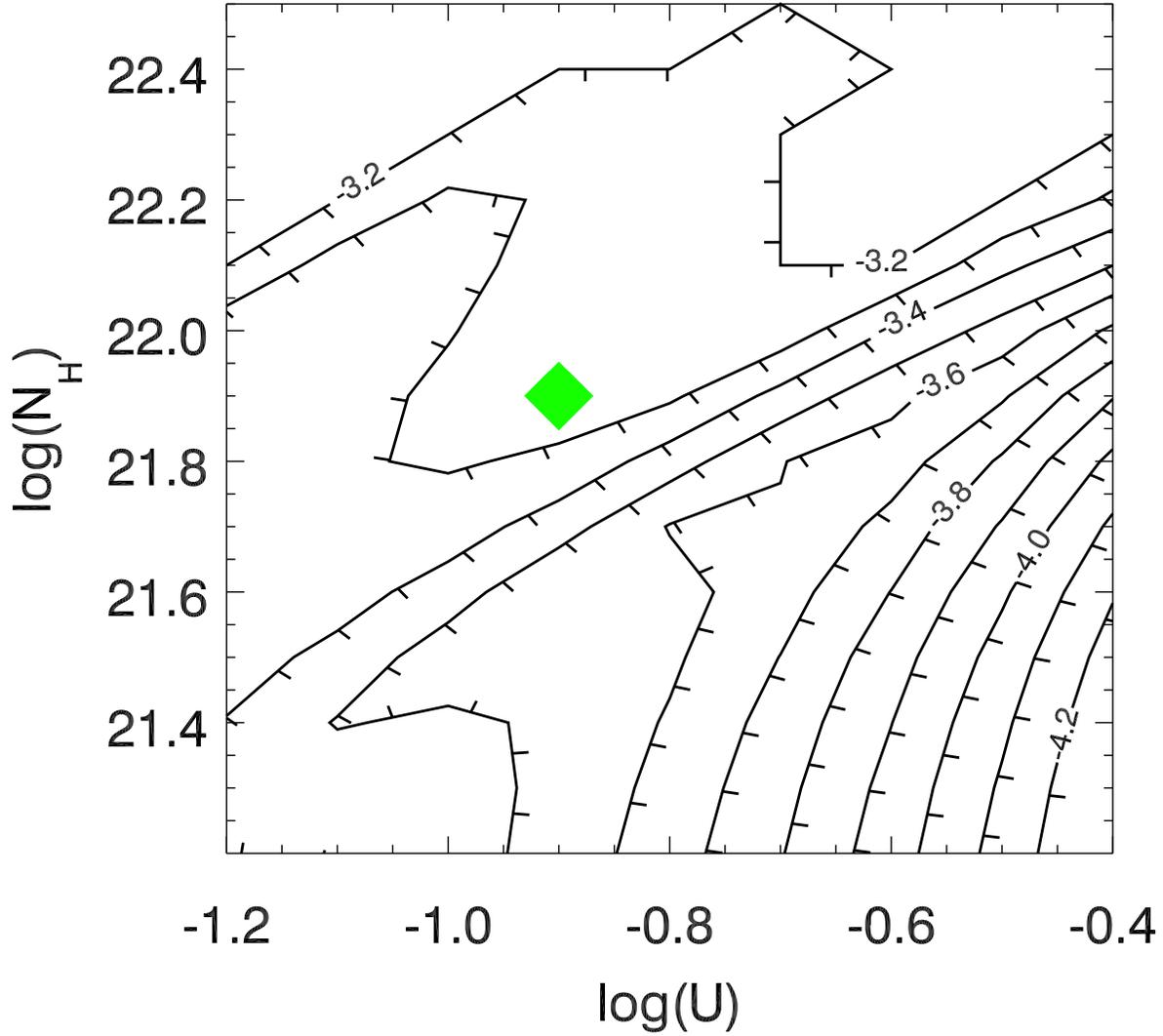}\\
 \caption{This plot shows contours of the logarithm of the predicted ratio of column density between \pv\ and \civ, $\log(N[$\pv$]/N[$\civ$])$, as a function
of the ionization parameter $U$ and the total hydrogen column density of the absorber $N_H$. Assuming solar abundances and the photionization solution
found in Section~\ref{physstate} (represented by a solid diamond), Cloudy models predicts a ratio of column density of $\log(N[$\pv$]/N[$\civ$]) \sim -3.3$.}
 \label{ioni_plot} 
\end{figure}

For outflows that show a significant \pv\ trough, our investigation gives us a unique empirical opportunity to contrast
the apparent and real optical depth of the \civ\ trough arising from the same outflow. As can be seen in Figure~\ref{alllineprof},
the residual intensity of the \civ\ trough in the deepest part of component 2 is $I_i \sim 0.025$, which yields an
apparent optical depth of $\ln(I_i) \sim 4$. The real optical depth of \civ\ at that velocity can be estimated using our derived
photoionization parameters for the absorber along with the knowledge of the column density and abundance of an unsaturated line
like \pv. Assuming solar abundances for simplicity (the abundance of phosphorus does not deviate widely from solar, see Section~\ref{phosab})
we represent in Figure~\ref{ioni_plot} the expected column density ratio $\log(N[$\pv$]/N[$\civ$])$ as a function of $U$ and $N_H$.
Using the ionization parameter of $\log(U) = -0.9$ and total hydrogen column density of $\log(N_H)=21.9$ found in Section~\ref{physstate} 
we find $\log(N[$\pv$]/N[$\civ$]) \sim -3.3$. The ratio of optical depth for the two transitions is then given by:
\begin{equation}
\frac{\tau[\mathrm{C} \textsc{iv} ]}{\tau[\mathrm{P} \textsc{v} ]}=\frac{N[\mathrm{C} \textsc{iv} ]}{N[\mathrm{P} \textsc{v} ]}
\frac{(f\lambda)_{\mathrm{C} \small{ \textsc{iv}} }}{(f\lambda)_{\mathrm{P} \small{\textsc{v}} }}=
10^{3.3}\frac{0.19}{0.45}\frac{1548.20}{1117.98} \sim 1200,
\label{tau}
\end{equation}
where $f$ is the oscillator strength and $\lambda$ is the wavelength
of the given transition. Using the partial covering method on the \pv\ doublet troughs we obtain a real
optical depth for \pv $\lambda$1117.98 of 2.7. Therefore, equation (\ref{tau}) yields a real \civ $\lambda$1548.20 optical
depth of $\sim$3200! Or a value almost 1000 times larger than the apparent optical
depth of \civ ($\sim 4$). Thus the detection of \pv\ troughs will indeed result in heavily saturated
\civ\ line profiles, with a real optical depth roughly a thousand times higher than the apparent one.

\section{Summary}
\label{conclu}

In this paper, we studied the UV outflow associated with the quasar SDSS J1512+1119 on the basis
of new, medium resolution VLT/X-Shooter data. The extended wavelength coverage of the instrument
allowed us to detect the outflow components in a multitude of ionic species. In particular we
report the detection of deep \pv\ absorption troughs in kinematic component 2 as well as the detection
of \siv\ and \siv*. A detailed analysis of the \siv* line profile allowed us to detect the weak
$\lambda 1073.51$ transition revealing a \siv\ column density larger than suggested by the
apparent depth of the absorption troughs of the $\lambda 1072.97$ transition.

Photoionization modeling of the absorber revealed that the absorber is thick, though the non-detection of significant \feii\ absorption troughs guarantees the absence of
a significant amount of \hi\ bound-free opacity.
Our accurate determination of the total \pv\, \siv\, \hei* and lower ionization species column densities allowed us to
characterize the physical state of the gas. We find that for the range of ionization parameters relevant for the present
absorber, the phosphorus abundance relative to helium is consistent with the solar value.
Using the parameter derived from the photoionization analysis,
we show, as suggested in \citet{Hamann98}, that a line such as the ubiquitous \civ\ is heavily saturated.
The \civ\ column density derived from the apparent depth of the absorption line profile underestimates the
column density by a factor of $\sim$1000, providing a very poor estimate of its 
true column density. 
 
The  phosphorus abundance we find is in disagreement with the extreme phosphorus abundances reported in the early literature \citep[e.g.][]{Junkkarinen95,Junkkarinen97,Hamann98}.
Other elemental abundances are found to be in agreement with the solar values. The fact that the abundances are similar to the solar values for an odd (\pv) and even (\siv)
element points to enrichment by relatively ``normal'' galactic stellar populations \citep[e.g.][]{Hamann97a} rather
than the more exotic mechanism proposed by \citet{Shields96} that would significantly enhance the P/S ratio \citep{Hamann98}.

\section*{ACKNOWLEDGMENTS}

B.B. would like to thank Pat Hall for suggesting looking at the \feiii* lines, Manuel Bautista for providing
critical densities for these lines and also Martino Romaniello and the ESO Back-end Operations Department
for pointing out the existence of ESO-Reflex. We thank the anonymous referee for valuable suggestions that
improved the paper as well as the suggestion of the use of a diagnostic plot similar to the one presented 
in Figure 8. We acknowledge support from NASA STScI grants GO 11686 and GO
12022 as well as NSF grant AST 0837880.


\bibliographystyle{apj}

\bibliography{astro}{}


\end{document}